**Potential curves illustrating a dissipative self-assembly system and the meaning of away-from-equilibrium**

Yoshiyuki Kageyama,* and Goro Maruta

Dissipative self-assembly is a recently proposed concept in the field of non-equilibrium systems. This concept catches the reader's attention owing to its relationship with life phenomena. Recently, Credi's group has reported the behavior of a directional supramolecular pump that works under continuous light irradiation with the regime of dissipative self-assembly.[1] Despite their remarkable results, the thermodynamic aspects illustrated in their paper have the potential to mislead readers about the chemistry of non-equilibrium systems.

A potential landscape explaining the autonomous molecular pump system, reported by Credi *et al.*[1], is presented in Figure 1a. The vertical axis depicts total Gibbs free energy and the horizontal axis represents the composition of the system. Therein, the potential curve has two minimum points: the lowest minimum point indicates the global thermodynamic minimum while the other point corresponds to a local equilibrium state or a kinetically trapped state. In addition, a dissipative state, which is described in their paper as a non-equilibrium steady state wherein externally supplied energy is consumed and transferred out of the system, is indicated on the slope of the curve. Similar figures without labels for horizontal axes also have been published by others who proposed dissipative self-assembly.[2-6] However, it is doubtful that the figures illustrate the state of the system correctly, which risks leading to the misunderstanding of non-equilibrium states of molecular systems.

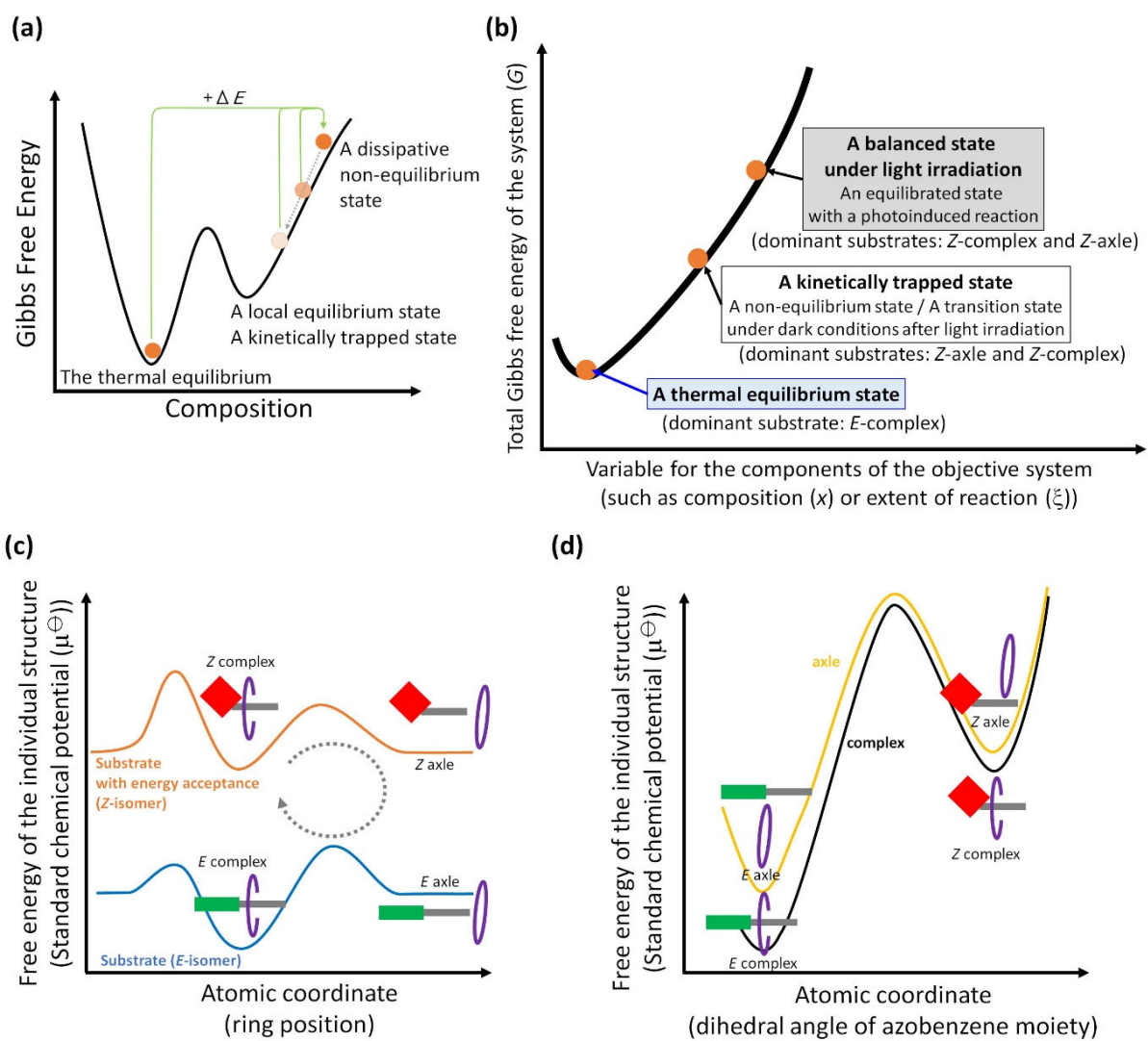

**Figure 1.** (a) Schematic illustration of the landscape of the total Gibbs free energy of a chemical system proposed by Credi *et al*. (The figure is quoted from lit. 1b with CC BY license and modified.) Similar figures have been provided by other researchers of dissipative self-assembly, but they are inaccurate. (b) A more accurate illustration of the landscape of the total Gibbs free energy of a chemical system in their study. (c) The standard chemical potential diagram with the ring position as the horizontal axis. (d) The standard chemical potential diagram with the dihedral angle of the azobenzene moiety as the horizontal axis. In their experiment, the molarity between *E*-complex, *Z*-complex, *Z*-axle, and *E*-axle shifts with changes in light intensity. Immediately after turning off the ultraviolet light, the system turns into a state composed mainly of *Z*-complex and *Z*-axle. In Credi's paper, this state is referred to as a kinetically trapped state as well as a local equilibrium state: in their system, the activation energy for *Z*-to-*E* thermal isomerization (d) is very large compared to the activation energy for the association–dissociation

reaction (c).

Herein, we consider the shape of the free energy landscape of the chemical system in which the horizontal axis (variable $x$) indicates the composition that changes along with the objective process in a system while the vertical axis is the total Gibbs free energy of the system ($G$). Composition is related to the concept of extent of reaction ($\xi$). At the minimum point of the Gibbs free energy potential curve, the change in free energy around the point (d$G$/d$x$) equals 0, meaning that the point represents a stable equilibrium point and that the involved chemical reactions are balanced. The state never shifts spontaneously from the minimum point because $\Delta G$ from this point is greater than zero. Next, we consider the kinetically trapped state. The term "kinetically trapped" communicates the notion that the expected reaction around the state is quite slow. Therefore, kinetically trapped states must exist on the slope of the potential curve rather than at the curve minimum, as shown in Figure 1b. According to Credi *et al.*, "the local equilibrium of the *Z* complex self-assembly reaction" is a kinetically trapped state, and they pointed out that the state will be relaxed to the more stable state by thermal *Z*-to-*E* isomerization, the velocity of which is much slower than the association–dissociation reaction. Thus, the kinetically trapped state in Credi's experimental system should be illustrated as shown in Figure 1b, rather than as in their figure, which is depicted in Figure 1a.

We should note that the Gibbs free energy of a chemical system is very different from the standard chemical potential ($\mu^{\ominus}$) of individual substances (the standard chemical potential for each structure of a couple of the molecules), as shown in Figure 1c–d. In the curve of the chemical potential versus the atomic coordinate, multiple valleys and peaks may be present. The total Gibbs free energy is the sum of the chemical potentials and considers the distribution of the components; therefore, there are generally no secondary minimums in the landscape of the total Gibbs free energy of a chemical system if the horizontal axis is reasonable. If there is a secondary minimum point on the potential curve, as shown in Figure 1a and other papers [2-6], the point is either an equilibrium point in another phase or is in another system (whose potential curve does not intersect with the potential curve of the original phase or system). Otherwise, the horizontal axis for the curve does not represent the chemical process in the objective system.

Next, we wish to address the definitions of equilibrium and away-from-equilibrium, which are frequently used in papers on dissipative self-assembly in chemical systems. According to the IUPAC Gold Book,[7] chemical equilibrium is explained as "reversible processes ultimately reach a point where the rates in both directions are identical, so that the system gives the appearance of having a static composition at which the Gibbs energy, $G$, is a minimum." It can be translated using the extent of the reaction ($\xi$) such that "$d\xi/dt$ is zero at which $dG/d\xi$ is zero (minimum)". This explanation is effective if the objective system is a closed system or an isolated system (Figure 2a). However, in an open system with energy flow, the fact that $dG/d\xi$ is zero (i.e., at minimum) does not correspond to the fact that $d\xi/dt$ is zero.[8] Therefore, two interpretations for "away-from-equilibrium" currently exist. One is the "$dG/d\xi \neq 0$" state, where $G$ is a potential value assumed for the system without energy flow. The detailed balance is broken in such a system, but macroscopic fluctuation in $\xi$ is not taken into consideration. The other interpretation is that the "$d\xi/dt \neq 0$" state differs from the balanced state; thus, $dG/d\xi$ is never zero. Credi and many other chemists who study dissipative self-assembly [2-6] or single molecule dynamics [9] employ the former interpretation, while Prigogine and other researchers follow the latter interpretation and have constructed their theory of dissipative structures accordingly [10,11] (Figure 2a). Prigogine has focused on two types of behavior in open systems with energy flow: the so-called non-equilibrium systems. One type is "linear non-equilibrium thermodynamics," whereby the system will achieve a balanced state ($d\xi/dt = 0$). Meanwhile, the other type is "non-linear non-equilibrium thermodynamics," whereby the reactions sustainably occur without being balanced; therefore, the composition of the system keeps fluctuating ($d\xi/dt \neq 0$) (Figure 2b). The term "far-from-equilibrium" has been employed for the latter behavior, and the generation of dissipative structures has been found in such systems. Examples include Bérnard convection and the Belousov–Zhabotinsky reaction in which the reaction components oscillate in the macroscopic part of the systems as well as recent studies of a self-oscillatory crystal[12] and self-oscillatory association and dissociation of amphiphiles[13,14]. Herein, we do not choose which of the two conflicting interpretations is more appropriate. However, as authors, reviewers, and readers, we should be aware that there are two different interpretations for "away-from-equilibrium" and related terms. Identical terms with two different definitions should not be used within a single paper. Likewise, the same benefit or utility from the phenomenon that each implies should not be expected. For example, in the introduction of Credi's paper, large-amplitude fluctuation in the composition and macroscopic motion were expected; however, they were not observed in the experimental results.

The terminology of dissipative self-assembly in chemistry was first summarized by Grzybowski *et al.* in 2006.[15] They introduce the collective dynamics of microtubules, bacteria, fish, metal beads, fluids, and gel particles as examples of "dissipative (dynamic) self-assembly" and show macroscopic and directional mechanical dynamics. It is noteworthy that the components are larger than synthesized molecules and are propulsive (whenever they are active or passive). In fact, the objects were originally far-from-equilibrium, even according to Prigogine's definition. On the other hand, the challenge of the currently proposed dissipative self-assembly concept in chemistry is the construction of collective and directional macroscopic dynamics ($d\xi/dt \neq 0$) from the very tiny behavior of molecular machines that work under conditions of $dG/d\xi \neq 0$. The principal question is whether chemical systems exhibiting macroscopic dynamics can be created by means other than Prigogine's dissipative structure theory or not. For this aim, directional molecular pumps and motors, such as the supramolecular complex in Credi's study, are very insightful agents.

In conclusion, the thermodynamic landscape was not illustrated accurately or unambiguously in recent papers on dissipative self-assembly by Credi *et al.*[1] and other groups. Additionally, the term "away-from-equilibrium" was used in a misleading manner. Therefore, we hope that the current paper will alert researchers in the field of dissipative self-assembly to these conceptual equivocations so that future publications will be more accurate.

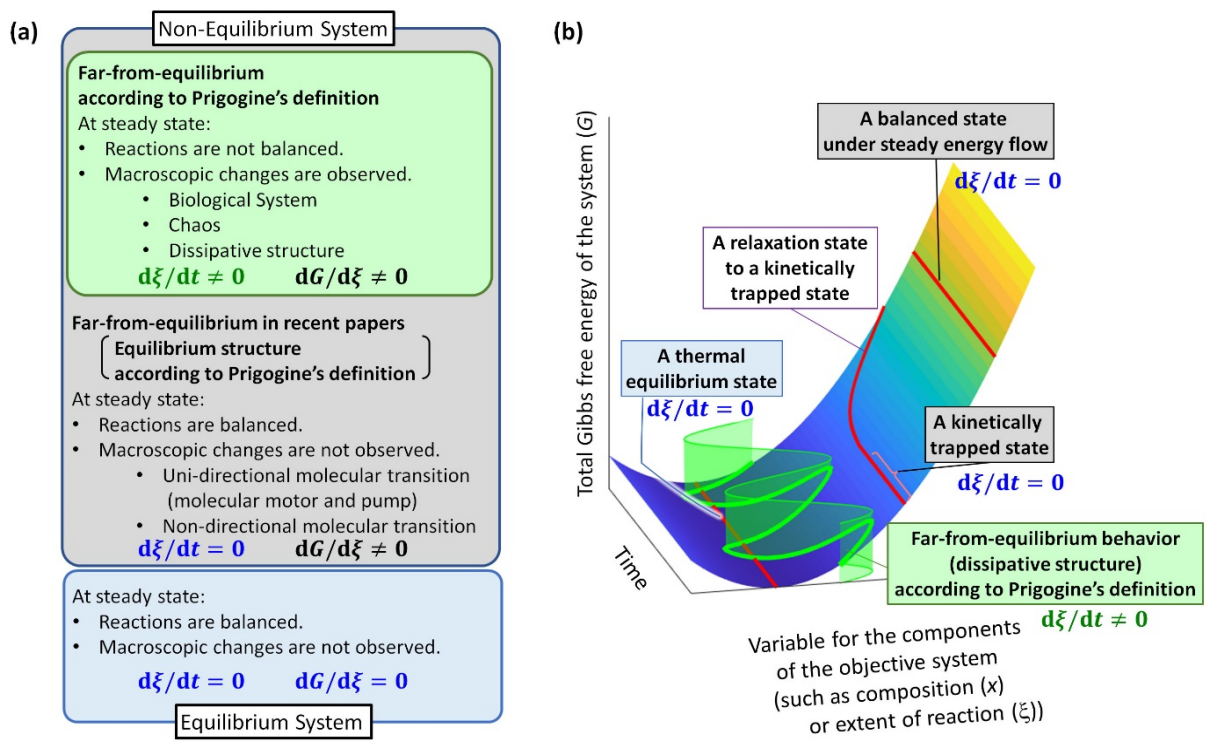

**Figure 2.** (a) Descriptions of non-equilibrium and equilibrium chemical systems according to different definitions. (b) Schematic illustration of a potential landscape including the time axis and thermodynamic states. The curve of the dissipative structure is drawn to show the fact that the composition is changing with time: Gibbs free energy surface may change with time in such a system.

1   (a) Corra, S. *et al.* Kinetic and energetic insights into the dissipative non-equilibrium operation of an autonomous light-powered supramolecular pump. *Nature Nanotechnology* **17**, 746-751 (2022). https://doi.org:10.1038/s41565-022-01151-y; (b) Corrà S, *et al.* Kinetic and energetic insights in the dissipative non-equilibrium operation of an autonomous light-powered supramolecular pump. *ChemRxiv. Cambridge: Cambridge Open Engage* (2021). https://doi.org/10.26434/chemrxiv-2021-8pwcc

2   Chen, X.-M. *et al.* Light-activated photodeformable supramolecular dissipative self-assemblies. *Nature Communications* **13**, 3216 (2022). https://doi.org/10.1038/s41467-022-30969-2

3   Grötsch, R. K. *et al.* Pathway Dependence in the Fuel-Driven Dissipative Self-Assembly of Nanoparticles. *Journal of the American Chemical Society* **141**, 9872-9878 (2019). https://doi.org:10.1021/jacs.9b02004

4   Sang, Y. & Liu, M. Hierarchical self-assembly into chiral nanostructures. *Chemical Science* **13**,

633-656 (2022). https://doi.org:10.1039/D1SC03561D

5	Sorrenti, A., Leira-Iglesias, J., Markvoort, A. J., de Greef, T. F. A. & Hermans, T. M. Non-equilibrium supramolecular polymerization. *Chemical Society Reviews* **46**, 5476-5490 (2017). https://doi.org:10.1039/C7CS00121E

6	Ragazzon, G. & Prins, L. J. Energy consumption in chemical fuel-driven self-assembly. *Nature Nanotechnology* **13**, 882-889 (2018). https://doi.org:10.1038/s41565-018-0250-8

7	IUPAC. Compendium of Chemical Terminology, 2nd ed. (the "Gold Book"). Compiled by A. D. McNaught and A. Wilkinson. Blackwell Scientific Publications, Oxford (1997). Online version (2019-) created by S. J. Chalk. ISBN 0-9678550-9-8. https://doi.org/10.1351/goldbook.

8	Astumian, R. D. Design principles for Brownian molecular machines: how to swim in molasses and walk in a hurricane. *Physical Chemistry Chemical Physics* **9**, 5067-5083 (2007). https://doi.org:10.1039/B708995C

9	Astumian, R. D. Microscopic reversibility as the organizing principle of molecular machines. *Nature Nanotechnology* **7**, 684-688 (2012). https://doi.org:10.1038/nnano.2012.188

10	Glansdorff, P. & Prigogine, I. *Thermodynamic theory of structure, stability and fluctuations*. (John Wiley & Sons Ltd 1971).

11	Nicolis, G. & Prigogine, I. *Self-Organization in Nonequilibrium Systems* (John Wiley & Sons, Inc., 1977).

12	Ikegami, T., Kageyama, Y., Obara, K. & Takeda, S. Dissipative and Autonomous Square-Wave Self-Oscillation of a Macroscopic Hybrid Self-Assembly under Continuous Light Irradiation. *Angewandte Chemie International Edition* **55**, 8239-8243 (2016). https://doi.org:https://doi.org/10.1002/anie.201600218

13	Leira-Iglesias, J., Tassoni, A., Adachi, T., Stich, M. & Hermans, T. M. Oscillations, travelling fronts and patterns in a supramolecular system. *Nature Nanotechnology* **13**, 1021-1027 (2018). https://doi.org:10.1038/s41565-018-0270-4

14	Howlett, M. G., Engwerda, A. H. J., Scanes, R. J. H. & Fletcher, S. P. An autonomously oscillating supramolecular self-replicator. *Nature Chemistry* **14**, 805-810 (2022). https://doi.org:10.1038/s41557-022-00949-6

15	Fialkowski, M. *et al.* Principles and Implementations of Dissipative (Dynamic) Self-Assembly.

**Acknowledgements**

Y.K. thanks Dr. Yutaka Sumino at Tokyo University of Science and Dr. Shoichi Toyabe at Tohoku University for fruitful discussions. Y.K. acknowledges the support of JSPS KAKENHI (Grant Number JP18H05423) in Scientific Research on Innovative Areas "Molecular Engine."

**Author contributions**

Y.K. mainly wrote the manuscript, and G.M pre-reviewed and assisted with the preparation of the manuscript.

**Competing interests**

The authors declare no competing interests.